\documentclass[aps,preprint,showpacs,preprintnumbers,amsmath,amssymb]{revtex4}

\usepackage{graphicx}
\begin{document}

\title{Ground state fragmentation of repulsive BEC in double-trap potentials}
\author{Alexej I.\ Streltsov}
\author{Lorenz S.\ Cederbaum} 
\affiliation{Theoretische Chemie,Universit\"at Heidelberg, D-69120 Heidelberg, Germany}
\author{Nimrod Moiseyev}
\affiliation{Department of Chemistry and Minerva Center of Nonlinear Physics in Complex Systems.
Technion-Israel Institut of Technology Haifa 32000,Israel}
\date{\today}

\begin{abstract}
The fragmentation of the ground state of a repulsive condensate immersed into a double-trap potential 
is found to be a general and critical phenomenon. 
It takes place for a given number of bosons 
if their scattering length is larger than some critical value 
or for a given value of the scattering length if the number of bosons is above some critical number. 
We demonstrate that the geometry of the inner trap 
determines these critical parameters while the number of the fragments 
and the fraction of bosons in the various fragments can be manipulated by the outer trap.
There is also a maximal number of bosons for which the ground state is fragmented.
If this number is exceeded, the fragmented state becomes a very low-lying excited state
of the condensate. This maximal number of bosons can be substantially manipulated by varying
the inner and outer traps.
To study three-fold fragmentation   
we have chosen a potential well with two barriers as the inner trap 
and embedded by two types of outer ones.
A many-fold fragmentation is also addressed.
\end{abstract}

\maketitle

\section{ Introduction }
%================= Condensation and Fragmentation ==========================
The general mathematical formulation of the condensation phenomenon for an ideal gas in equilibrium
has been given by Penrose and Onsager \cite{Penrose} in 1956.
They considered a system of N interacting bosons and its reduced one-particle density matrix.
The eigenvalues of this matrix are called {\it occupation numbers} and the
eigenvectors are referred to as {\it natural orbitals}.
If an ideal gas of free bosons forms a BEC, then only a single natural orbital is macroscopically occupied.
The extension of this concept on to several macroscopically occupied orbitals
is the basis of the fragmentation phenomenon. 
The BEC is called {\it fragmented} if several natural orbitals have macroscopic occupation numbers \cite{Nozieres}.
Although these original definitions have been formulated in the thermodynamic limit $N\to\infty$,
the reduced one-particle density matrix can also be applied to study condensation \cite{NCondensation} and 
fragmentation \cite{NFragmentation} in finite-N bosonic systems.

The first measurements of an interference between two expanding and overlapping condensates 
\cite{Interference} have stimulated a great interest in the theoretical studies of fragmentation.
An initial quantum state of such a system is supposed to be two-fold fragmented i.e.
two spatially separated orbitals are macroscopically occupied. The theoretical studies on
relative phases \cite{phase1,phase2,phase3} between the fragments and related questions on 
dynamical evolutions of this state \cite{JJ,JJ1} have been a subject of numerous discussions. 
A natural extension of these effects to an array of multiple wells with a many-fold fragmented initial state
\cite{phaseM} has been initiated by the recent experiments 
in optical \cite{OpttrapExper} and hybrid traps \cite{HybridTrap}.
% have extended those problems naturally to an array of multiple wells and a many-fold fragmentation \cite{phaseM}. 

In contrast, the fragmentation of the ground state of BEC has been predicted only for 
a few systems where it is enforced by the spatial or spin symmetry of the systems.
%====================== Fragmentation of Atractive BEC GS symm.brk.states======================
In particular, the ground state of an attractive condensate in a perfect ring 
is found to be fragmented \cite{UedaPRA}.
%The resent investigation of this phenomenon at the level of configuration 
%interaction ~\cite{Ofir} demonstrates that fragmentation takes place for any 
%non-zero attractive interaction and for any number of particles.
The rotational symmetry of the corresponding Hamiltonian permits as natural orbitals only plane waves ~\cite{Ofir}. 
The many-body wave-function of this system starts to differ from a function composed by 
plane-waves for any 
non-zero attractive interaction and the respective one-particle density matrix having
plane-waves as natural orbitals must exhibit several non-zero eigenvalues. 
This eventually leads to fragmentation %\cite{Nozieres} 
and its origin is spatial (rotational) symmetry.
Another example is an attractive BEC in a symmetric double-well potential where 
the ideal symmetry of the potential ~\cite{BMF} causes the fragmentation of the ground state.
The stability of all these fragmented states with respect to a small asymmetric perturbation is 
still an open question. There are indications that these symmetry related fragmentations
disappear  upon symmetry breaking.
%====================== Fragmentation of Spinor BEC GS symm.brk.states======================
Fragmentation can also take place if bosons have additional internal degrees of freedom. 
An example of such a system is a system made of Bose particles with non-zero spin, a so-called spinor condensate.
In spinor condensates different spin components may have different spatial extensions 
in the presence of an external magnetic field. 
Indeed, the ground state of Spin-1 Bose gas in a uniform magnetic field was proved to be fragmented ~\cite{HOYip}.

%=============================== Description of fragmentation ===================================
The single condensate picture and its mean-field description via 
the Gross-Pitaevskii (GP) equation has been a very successful approximation 
and can explain many experiments, see, e.g., Refs.~\cite{review1,review2} and references therein.
However, this mean-field is incapable by definition to describe fragmentation, 
since only one orbital is involved.
Recently, a more flexible mean-field approach allowing for bosons to reside in 
different orthonormal one-particle functions has been formulated ~\cite{BMF}.
This intrinsic ability to describe fragmentation makes it very attractive for theoretical investigations
and predictions. In the framework of this best mean-field (BMF) it is possible to answer the question
whether fragmentation is energetically favorable or not.
In particular, a repulsive BEC in an asymmetric double-well potential can be in a stable two-fold fragmented 
state ~\cite{BMFES}, but the energy of this state
is higher than the energy of the respective non-fragmented ground
state of the condensate, i.e., it is an excited state of the BEC.
This result is in agreement with general predictions derived by Nozieres \cite{Nozieres}
that interaction prevents fragmentation in repulsive condensates.
We shall demonstrate below, however, that in appropriate traps, repulsive BECs can exhibit fragmentation
in the ground state.
%1)NO Fragmentation for GS of attractive HO potential  O.Elgaroy and C.J.Pethick PRA,v59,1711

% =============================== Finite-N infinite-N limits    ======================
%Another very interesting property of this approach is an ability to
%distinguish the system of different numbers of bosons.
% =============================== Organization of the article ======================
In the present paper we use the best mean-field approach to study the ground state of a system of N 
identical bosons with positive scattering length immersed into a double-trap external potential.
A double-trap potential consists of an inner trap embedded in a wider outer trap.
The paper is organized as follows. In Sec.II we briefly discuss the three-orbital 
best mean-field formalism. We also provide a very transparent example which illustrates that 
a finite number of particles and well-separated multiple potential wells 
are favorable conditions for fragmentation.
In Sec.III. we propose specific shapes of trap potentials and discuss why 
fragmentation is expected.
We demonstrate in Sec.IV that indeed in the ground state macroscopic occupation of three 
single-particle functions is energetically more favorable than accumulation of all particles
in a single orbital. We also show that fragmentation is a general phenomenon which, depending on the 
trap potentials, may take place for any number of particles.
In Sec.V we demonstrate that the number of fragments, the shapes of the respective one-particle orbitals 
and their occupation numbers can be manipulated by the proper choice of the outer trap.
Here, we also verify that fragmentation of the ground state is a critical phenomenon, because it occurs  
when the number of bosons exceeds some critical number at a fixed scattering length,
or at some critical scattering length if the number of bosons is fixed.
The interplay between critical parameters of the fragmentation and the geometry of the inner trap
forms the content of Sec.VI. 
A discussion of the factors suppressing fragmentation in the ground state is given in Sec.VII.
Next, we briefly address many-fold fragmentation in a multiple wells in Sec.VIII.   
Finally, Sec.IX summarizes our results and conclusions.

\section{Mean-Field Theories}
%=============Hamiltonian =============================================
We consider a system of N identical bosons interacting via a $\delta$-function contact potential
$ W(\vec{r}_i-\vec{r}_j)=\lambda_0\,\delta(\vec{r}_i-\vec{r}_j) $, where $\vec{r}_i$ is the position of the i-th
boson and the nonlinear parameter $\lambda_0$ is related to the s-wave scattering 
length of the bosons \cite{review2}.
%Introducing the field operator $\Psi$, the Hamiltonian of this system is
%\begin{equation}
%H=\int d\vec{r} \Psi^{\dagger} h\,\Psi + \frac{\lambda_0}{2}\int d
%\vec{r}\,\Psi^{\dagger}\,\Psi^{\dagger}\,\Psi\,\Psi
%\nonumber
%\end{equation}
%where $h(\vec{r})=\hat{T}\,+\,\hat{V}(\vec{r})$ is the unperturbed one-particle Hamiltonian consisting of the
%kinetic operator $ \hat{T} $ and the external potential $\hat{V}(\vec{r})$.

%=============Standart One -Orbital Mean-Field Description  =============================================
The standart mean-field description of the interacting system is obtained by assuming the ground state
wave function $\Psi$ to be a product of identical spatial orbitals $\varphi$:
$\Psi(\vec{r}_1,\vec{r}_2,\ldots,\vec{r}_N)= \varphi(\vec{r}_1)\varphi(\vec{r}_2)\cdots\varphi(\vec{r}_N)$.
The energy $E\,\equiv\,<\Psi|\hat{H}|\Psi>$, defined as the expectation value of the $\hat{H}$, reads
\begin{equation}
E_{GP}=N\{\int\varphi^* h\,\varphi\,d\vec{r}+\frac{\lambda}{2}\int|\varphi|^4\,d\vec{r}\},
\label{GP_energy}
\end{equation}
where $\lambda=\lambda_0(N-1)$ is the interaction parameter and 
$h(\vec{r})=\hat{T}\,+\,\hat{V}(\vec{r})$ is the one-particle Hamiltonian consisting of the
kinetic operator $ \hat{T} $ and the external potential $\hat{V}(\vec{r})$.
The mean-field equation for bosons residing in a single orbital obtained by minimizing
this energy is the well-known Gross-Pitaevskii equation \cite{GP1,GP2}.
\begin{equation}
\{\:h(\vec{r})+\lambda_0(N-1)|\varphi(\vec{r})|^2\}\,\varphi(\vec{r})=\,\mu_{GP}\,\varphi(\vec{r}).
\label{GP_orbital}
\end{equation}
By definition, this equation cannot describe fragmentation because all bosons reside in a
single orbital. The reduced one-particle density operator and the corresponding spatial density 
are given by  
$\rho_{GP}(\vec{r},\vec{r'})=\varphi^*(\vec{r'})\varphi(\vec{r})$ and 
$\rho_{GP}(\vec{r})=|\varphi(\vec{r})|^2$ respectively.

%=============Extended  N-Orbital Mean-Field Description  =============================================
In order to describe the fragmentation on a mean-field level, a
more general ansatz for wave-function must be used. 
In the present study three orbitals $\phi_1$, $\phi_2$ and $\phi_3$ 
with particle occupations $n_1$,$n_2$, $n_3$ and $n_1+n_2+n_3=N$ are utilized. 
It will become evident below why three orbitals have to be used in the present investigation.
Since the details of the general derivation have already been published elsewhere ~\cite{BMF,BMFES},
we outline here only the main steps. With this ansatz the wavefunction now reads
\begin{eqnarray}
\Psi(\vec{r}_1,\ldots,\vec{r}_N)=
\hat{\cal S}\phi_1(\vec{r}_1)\cdots\phi_1(\vec{r}_{n_1})\phi_2(\vec{r}_{n_1+1})
\cdots\phi_2(\vec{r}_{n_2})\phi_3(\vec{r}_{n_2+1})
\cdots\phi_3(\vec{r}_{n_1+n_2+n_3}),
\end{eqnarray}
where $\hat{\cal S} $ is the symmetrizing operator.
The energy expression takes the form:
\begin{eqnarray}
 E = 
n_1 h_{11} + \lambda_0 \frac{n_1(n_1-1)}{2}\int|\phi_1|^4 d\vec{r}+ 
n_2 h_{22} + \lambda_0 \frac{n_2(n_2-1)}{2}\int|\phi_2|^4 d\vec{r}+ & & \nonumber\\
n_3 h_{33} + \lambda_0 \frac{n_3(n_3-1)}{2}\int|\phi_3|^4 d\vec{r}+ 
+2 \lambda_0 n_1 n_2 \int|\phi_1|^2 |\phi_2|^2 d\vec{r} + & & \nonumber \\
+2 \lambda_0 n_1 n_3 \int|\phi_1|^2 |\phi_3|^2 d\vec{r} + 
+2 \lambda_0 n_2 n_3 \int|\phi_2|^2 |\phi_3|^2 d\vec{r} 
\label{BMF_energy}
\end{eqnarray}
%It contains terms 
By minimizing  this energy with respect to the orbitals under the constraints
that they are orthogonal and normalized, i.e., $ <\phi_i|\phi_j>=\delta_{ij}$,
we get the following three coupled equations for the optimal orbitals:
\begin{eqnarray}
\{\:h(\vec{r})+\lambda_0 (n_1-1)|\phi_1(\vec{r})|^2+2 \lambda_0 n_2|\phi_2(\vec{r})|^2  
+2 \lambda_0 n_3|\phi_3(\vec{r})|^2 
\}\,
\phi_1(\vec{r}) =  \nonumber & & \\
= \mu_{11}\,\phi_1(\vec{r})+\mu_{12}\,\phi_2(\vec{r}) +\mu_{13}\,\phi_3(\vec{r}) \nonumber & & \\
\{\:h(\vec{r})+\lambda_0 (n_2-1)|\phi_2(\vec{r})|^2+2 \lambda_0 n_1|\phi_1(\vec{r})|^2 
+2 \lambda_0 n_3|\phi_3(\vec{r})|^2 
\}\,
\phi_2(\vec{r}) = \nonumber & & \\
= \mu_{21}\,\phi_1(\vec{r}) + \mu_{22}\,\phi_2(\vec{r})+\mu_{23}\,\phi_3(\vec{r}) \nonumber & & \\
\{\:h(\vec{r})+\lambda_0 (n_3-1)|\phi_3(\vec{r})|^2+2 \lambda_0 n_1|\phi_1(\vec{r})|^2 
+2 \lambda_0 n_2|\phi_2(\vec{r})|^2 
\}\,
\phi_3(\vec{r}) = \nonumber & & \\
= \mu_{31}\,\phi_1(\vec{r}) +\mu_{32}\,\phi_2(\vec{r}) + \mu_{33}\,\phi_3(\vec{r}).& &
\label{BMF_orbital}
\end{eqnarray}

%================== How good BMF is =======================================
Several key features of this approach should be mentioned.
First, this mean-field includes GP as a special case:
when the occupation of two of the three orbitals vanish, i.e. $n_2=n_3=0$, the system of equations
(\ref{BMF_orbital}) is reduced to the single equation (\ref{GP_orbital}) 
and the respective energy in Eq.\ref{BMF_energy}
coincides with the GP one (compare Eq.\ref{BMF_energy} and Eq.\ref{GP_energy}). 
Second, by construction, this method can describe fragmentation for a finite number of 
fragments. Indeed, in this three-orbital case the reduced one-particle density operator
can be written as
$\rho_{BMF}(\vec{r},\vec{r'})=\sum_{i}^3 \phi^*_i(\vec{r'})\phi_i(\vec{r}) n_i$ and 
the corresponding spatial density 
becomes $\rho_{BMF}(\vec{r})=n_1|\phi_1(\vec{r})|^2+n_2|\phi_2(\vec{r})|^2+n_3|\phi_3(\vec{r})|^2$. 
Third, the occupation number $n_i$ of all fragments are variational parameters, minimizing the full energy.
%=============== Variational parameters and how to get the minimuma (if it exist)
%Occupation $n_i$ are variational parameter for BMF, i.e.,
In order to find this optimal value of the energy, 
%, one should search for its optimal value (including zero occupations) which makes the energy stationary.
%Technically, this means that 
Eqs.\ref{BMF_orbital} are solved for all possible occupation numbers, 
and the respective energies, Eq.\ref{BMF_energy}, are evaluated and compared. 
As we discussed before, the results obtained for the specific occupation numbers $n_2=n_3=0$ 
are identical to the standart GP ones. Therefore, within this computational scheme  
we automatically clarify the question on the favorability of fragmentation.
%===================== Explaination of paramtrization ========================
% Finite N-questions
%This choice of parameters we will use further as well because
Fourthly, $\lambda=\lambda_0(N-1)$  is the only parameter involved in 
the GP energy per particle expression (see Eq.\ref{GP_energy}). Therefore,
the one-orbital mean-field cannot distinguish energetically between two system made of different numbers of bosons 
if they are characterized by the same $\lambda$.
In contrast to that, BMF treats these systems differently.
For example, at the GP mean-field level two systems made of 11 bosons with $\lambda_0=0.1$
and of 1000001 bosons with $\lambda_0=0.000001$ are characterized by the same energy per particle.
At the BMF level of description, if one of the systems is fragmented, 
the respective energies are different due to different fractional occupation numbers.
This observation defines $\lambda,N$ as a very natural choice of the parameters to study fragmentation at the 
general mean-field level. If fragmentation takes place, then the number of particles becomes a relevant parameter.
In the following, to compare systems made of a different number of bosons 
we adjust their $\lambda_0$'s in such a way that $\lambda$ is the same for each system.
Then, for a given trap potential the GP ansatz gives the same energy per particle for all these systems.

%=============== Why GP is insufficient  - and other mean-fields are needed
Large numbers of bosons and well-separated quantum levels of the single-well trap potential provided \cite{review2}
a justification of the single-orbital mean-field description of BEC.
However, recent experiments on optical trapping of BEC 
have initiated interest in studies of multi-well systems and questioned the validity of the one-orbital
mean-field description.
The ground and lowest excited states of a multi-well trap potential can be
almost degenerate and this opens a competition among the involved single-particle levels
and the degree of their occupations.
Therefore, the one-orbital mean-field description may be insufficient.
%============================ Optical lattices ====================================
%============================ Multiwell system ===================================
Another difficulty for the one-orbital mean-field arises when the number of wells 
is comparable to the number of particles.
In this case the average occupation number of each well (so called filling factor per lattice site ) 
can be of the order of several atoms.
%====================== Transparent example on three wells where BMF3 is needed================
%====================== Analisys of Three-Separated wells for finite number of N ==================
It is worthwhile to demonstrate that for such a situation 
a many-orbital mean-field is the best mean-field, since it
is energetically and physically more favorable than the one-orbital ansatz.

Let us consider a system of 3 repulsive bosons trapped in three equivalent, infinitely separated wells
which we denote left(l), central(c) and right(r).
Without loss of generality, we can assume that the lowest total energy is obtained if 
each well contains only one particle, i.e.,  $n_l=n_c=n_r=1$.
Since the wells are equivalent and infinitely separated, the wave-functions (orbitals) of each boson
$\psi_l,\psi_c,\psi_r$ are the same but localized at the different wells.
This implies also zero overlap between each pair of orbitals.
In other words, we have a system of three non-interacting bosons.
The GP orbital now reads $\varphi=(\psi_l+\psi_c +\psi_r)/\sqrt{3}$
while the BMF orbitals are the three orbitals $\phi_1=\psi_l,\phi_2=\psi_c$ and $\phi_3=\psi_r$.
A substitution of these functions into Eq.\ref{GP_energy} and Eq.\ref{BMF_energy} gives: 
\begin{eqnarray}
E_{1-orb}&=&h_{ll}+h_{cc}+h_{rr}+\frac{3(3-1) \lambda_0}{2}\int
(\frac{1}{9}|\psi_l|^4+\frac{1}{9}|\psi_c|^4+\frac{1}{9}|\psi_r|^4)d{\vec{r}} \nonumber \\
E_{3-orb}&=&h_{ll}+h_{cc}+h_{rr} \nonumber  
\end{eqnarray}
Inspection of these energies clearly shows that for repulsive interaction $\lambda_0>0$ the three-orbital 
description is energetically more favorable than the one-orbital (GP) approach.
The total energy of three equivalent non-interacting bosons is expected to be a 
sum of the energies of each particle. Therefore, only the BMF describes 
the physics correctly, while the GP energy contains an artificial term which can be 
considered as an interaction between the actually non-interacting subsystems.
For this example of three infinitely separated wells, the energy provided by the many-orbital mean-field (BMF) 
is lower than the GP one for any finite number of particles.
At the limit of very large N the energy difference between GP and BMF vanishes %becomes negligible
while the physics in the GP case still remains wrong.
For this trivial case the three-orbital ansatz is evident. However, the question whether the number of the 
fragments is always equal to the number of wells, 
deserves a more detailed investigation and will be reported elsewhere \cite{FiniteN1-2-3foldfrag}. 
%Conclusion: One-orbital mean-field description of BEC appears to be valid only at the large-N limit.

We close this section with a general remark concerning BMF.
For a given number of orbitals $m_0$, three in the present case, the BMF approach determines their
optimal occupation numbers minimizing the energy functional in Eq.\ref{BMF_energy}.
We call the result BMF($m_0$). As mentioned above, the calculation may provide the result that
some of these orbitals are not occupied, i.e., their occupation number is equal to zero. This, of course,
implies that the overall best mean-field is achieved with less than $m_0$ orbitals.
Generally, we arrive at the overall best mean-field if inclusion of more orbitals does not improve 
the description. In the present study $m_0=3$ and we have arguments that this choice leads to the overall best 
mean-field.

\section{Proposed Double-Trap potential}
Our proposed 1D double-trap potentials is shown in Fig.\ref{fig1}.
Effectively such a trap may be obtained as a superposition of two potentials (inner and outer).
We model the inner potential as:
\begin{equation}
V_{inner}(x)= \omega (\frac{x^2}{2}-A)e^{(-Bx^2)}
\label{InnerTrap}
\end{equation}
where $A$ and $B$ are parameters of the inner trap.
As an outer trap embedding the inner one we used either an infinitely deep square potential well
(infinite square well) with half-width equal $C$ or a smooth power potential $V_{outer}(x)=(0.035x)^{10}$.
The infinite square well is obtained by placing the infinite walls at $[-(C+\delta):+(C-\delta)]$. 
We introduced a small asymmetry parameter $\delta=0.01\pi$ to destroy the exact symmetry of the trap potential
in order to get rid of the effects of symmetry. 
In the following, as a default outer trap we use a square well with walls at $C_0=9.5\pi$ and 
$A_0=0.8,B_0=0.1$ as reference parameters of the inner trap.
The corresponding kinetic energy reads 
$\hat{T}=-\frac{\omega}{2}\frac{\partial^2}{\partial x^2}$ implying that coordinate $x$ 
is dimensionless and all energies and $\lambda_0$ are now in units of the frequency $\omega$.
%As an alternative to infinite walls quasi-"harmonic" potential $(0.035*x)^{10}$ may be used. 
%The periodic boundary conditions are equivalent to the BEC in a modulated Ring problem,
%however, the topology of the resulted potential ($V_{inner}+V_{outer}$) for this case
%differs from previous two.
%The resulted potential for the infinite squared well and quasi-harmonic potentials  has   
%three wells, while for a ring only two. 
%Moreover, for a Ring a periodic version of $V_inner(x)$ must be used and choose
%$R_{ring}=C_0/\pi$ as a ring radius in order to provide the numerical comparison of
%all considered potentials.
%================= END Examples of potentials =====================================================
%Here, for transparency of presentation we confine ourselves in the following to infinite walls as a default 
%outer trap potential.

%=================== Discussion with respect to Tunneling ==================================
An inner trap potential of the form given in Eq.\ref{InnerTrap} has been originally proposed 
to study BEC tunneling \cite{Nimrod0,Nimrod}. For the above reference parameters and one particle
this potential has only a single bound state and a set of metastable states at positive energies 
(so called resonances).
In the limit of non-interacting particles, all bosons will occupy this bound orbital, while 
for a non-zero repulsive interaction the bosons trapped inside the well may "flow out".
In this case, a competition between the bound state localized inside the well 
and the continuum states outside the barriers has been predicted \cite{Nimrod}. 
The experimental observation of the continuum outgoing waves is a delicate problem.
By placing a secondary trap potential beyond the barriers the continuum outgoing wave functions are 
discretized   and "transformed" to real functions which can be occupied by bosons and observed experimentally. 
%================= Discussoions on the specific form of secondary trap potential ========================
We shall demonstrate later that the specific shape of the secondary trap potential 
is rather of minor importance, while its width is a major factor.
The secondary trap potential should have a width capable to accumulate particles. 
The simplest choice is to place two infinite walls at some distance from the origin.
%, that equivalent to infinite squared well.
The fixed infinite walls can be replaced by some external embedding potentials with a smooth profile.
From an experimental point of view this means that the fragmentation phenomenon
may be observed in outer ordinary traps.
%================= Examples of Outer potentials =====================================================

\section{Fragmentation}
The combined potential ($V_{inner}+V_{outer}$) has three well-separated wells, see Fig.\ref{fig1}. 
Therefore, if fragmentation takes place, bosons will be accumulated 
in each of these three wells. 
More precisely, the reduced one-particle density of the system of N identical bosons
in this double-trap potential would have three {\it macroscopic} (with respect to N) eigenvalues.  
The condition that all three wells are well separated from each other implies that the respective 
eigenvectors (natural orbital) will be predominantly localized in each of these wells.

Indeed, within the framework of BMF(3) the ground state of the system of N bosons becomes three-fold fragmented.
In Fig.\ref{fig2} we present two sets of the orthonormal BMF orbitals and the respective densities corresponding 
to the system of $N=25$ and $N=6000$ bosons and compare them with the GP results. 
The interaction strengths of these systems have been chosen to keep the quantity $\lambda=\lambda_0(N-1)=1.3$ 
fixed for both systems as explained above. BMF(3) predicts fragmentation of the ground state 
as it provides a lower energy than GP does.

%======================== Density and energies discussion ===========================================
From Fig.\ref{fig2} one can see that the spatial densities 
(not to be confused with the reduced one-particle density) 
of GP, $\rho_{GP}=|\varphi|^2$, and of the three-orbital BMF, 
$\rho_{BMF}=n_1|\phi_1|^2+n_2|\phi_2|^2+n_3|\phi_3|^2$, are rather similar. 
At the same time the energy per particle provided by the BMF is lower than 
the respective GP one. 
%When the number of bosons is increased the BMF energy  approaches the GP energy from below.
Despite the fact that 
%for very large N's 
the spatial densities provided by
GP and BMF are similar, there is a substantial physical difference between the methods.
At the 1-orbital level of description the systems are unfragmented, while at the 3-orbital one they are
three-fold fragmented. 
%More detailed investigation on these effects will be published elsewhere \cite{FiniteN1-2-3foldfrag}.

%================== Here we fix and explain  N2~=N3=(1-N1)/2 and Fractional occupation numbers
By solving Eqs.\ref{BMF_orbital} for different occupation patterns we obtained 
the optimal fractional occupation numbers for the systems of $N=6000$  and  $N=25$ bosons.
For the sake of convenience we will from now on also use the term fractional 
occupation number for $n_i/N$  and express this quantity in $\%$ ($\frac{n_i}{N}100\%$).
These are found to be similar for both systems: 
$n_1/N\approx 68.58\%$ for the orbital localized in the central well 
and $n_2/N\approx n_3/N=17.71\%$ for the orbitals localized in the outer wells for $N=6000$
and $n_1/N\approx 67.5\%$  and $n_2/N\approx n_3/N=16.25 \%$ for $N=25$.
The fact that the occupation numbers of the orbitals localized in the left and right wells are approximately 
the same is explained by the slight asymmetry of the double-trap potential.
This observation allowed us to simplify the numerical search for the optimal values of the occupation numbers.
Instead of searching for the minimum of a functional $E(n_1,n_2,n_3)$ 
of two independent variables $(n_3\equiv N-n_1-n_2)$, we can start the search 
using $n_2=n_3$ and then relax this condition.
%In the following we will use the occupation of the inner well $N_1$ as this independent variable,
%and keep the occupation of the left and right equal, namely  $N2\approx N3=(N-N1)/2$.

It is convenient to use the fractional occupation $n_1/N$ of the orbital localized in the 
central well as a characteristic parameter of fragmentation.
If the fractional occupation $n_1/N=100\%$ then there is no fragmentation at all, while
for any other values of $n_1/N$ the fraction of bosons accumulating in the outer wells 
is defined as $100\%-n_1/N$. In the following we call the latter quantity the {\it fragmented} fraction
in the outer wells or briefly the fragmented fraction. 
For the specific examples depicted in Fig.\ref{fig2} ($\lambda=1.3$),
the fragmented fractions of the systems with $N=25$ and $N=6000$ bosons 
are $32.5\%$ and $31.42\%$, respectively.

This observation allows us to conclude that if the ground sates of the system made of 
a large number of bosons is fragmented, then   
any other system of bosons characterized by the same $\lambda$ and made of a smaller number of 
particles (of course, $N>1$) is also fragmented. The opposite does not apply, however.
For a given value of $\lambda$ there is a maximal number of bosons for which the ground state
is fragmented. This number depends on the trap potentials used and can be manipulated by
changing these potentials. For a discussion of this issue, see Sec.VII.

In the following study we confine ourselves to the system of $N=25$ particles, keeping in mind 
that for a larger number of particles the occupation numbers may differ within less than $~5\%$
as long as the condensate is fragmented (this has been verified numerically).

\section{Manipulating Fragmentation by varying the outer trap}
By "manipulation of the fragmentation" we mean the possibility to choose the shape of the trap potential
as well as the number of bosons (and possibly also their scattering length)
in such a way that all fragments acquire the desired occupation numbers.

As we briefly mentioned above, the bosons trapped by the inner potential alone in the absence
of the outer trap occupy the bound state (localized in the central well) as long as  $\lambda<\lambda_{cr}$.
Any change in the number of particles or in the scattering length such that
$\lambda=\lambda_0(N-1)$ becomes larger than $\lambda_{cr}$, imidiatelly initiates 
tunneling - the flow of bosons out of the central well \cite{Nimrod}.
If infinite walls are placed beyond the barriers, the system becomes closed 
and bosons are collected in the outer wells.
If all three wells are macroscopically occupied the system is fragmented.
%================== Questions arised ===================================================

Several questions arise in the presence of the outer trap.
Does the fragmentation phenomenon exist for any $\lambda$  
or is it characterized by some critical parameters, similarly to tunneling in the open system?
The second question may be formulated as follows:
Do the fractional occupation numbers depend upon  
the positions $\pm C$ of the walls of the outer trap and on the particular shape of this trap?

\begin{figure}
\includegraphics[width=11.2cm, angle=-90]{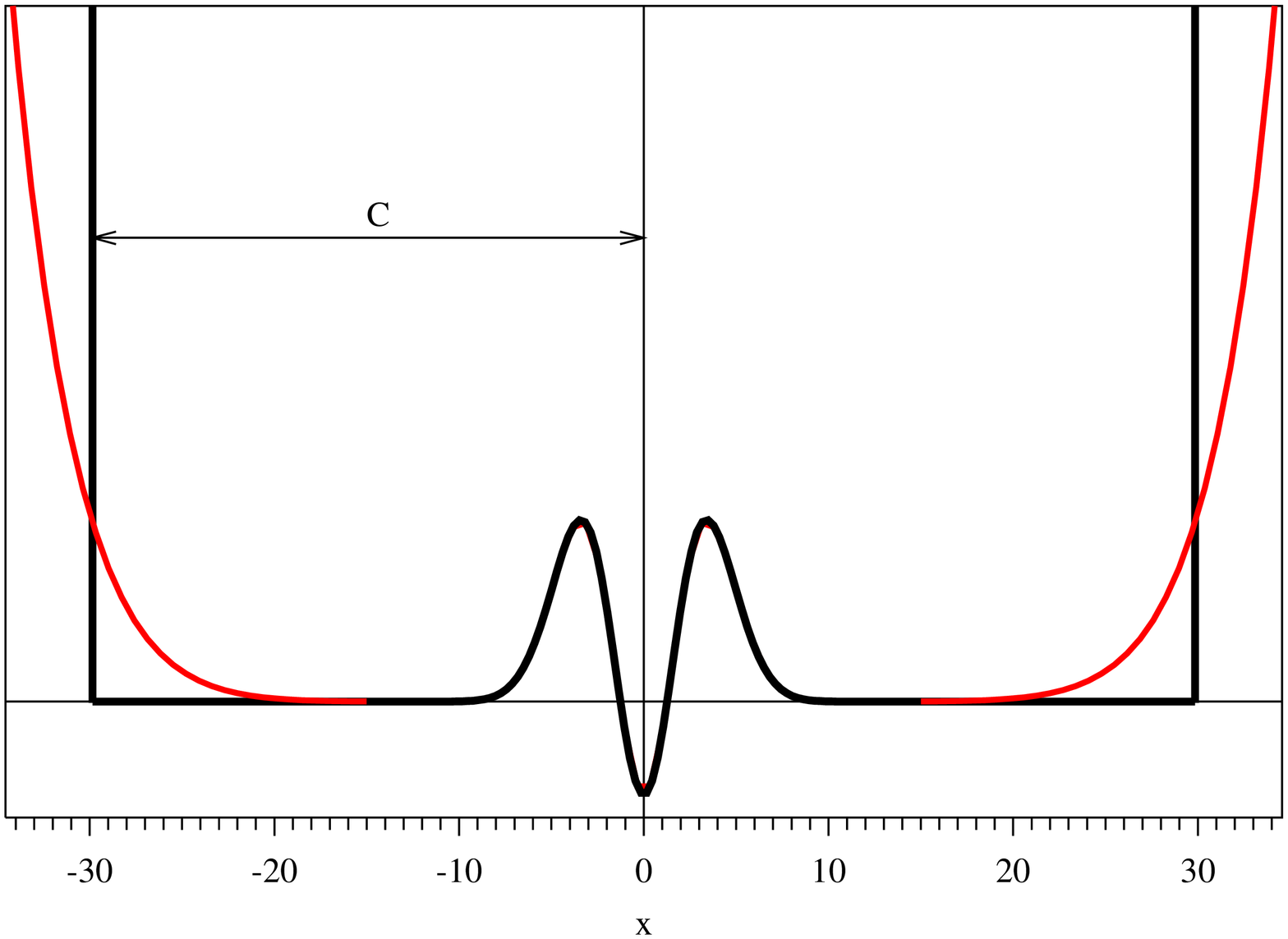}
\caption{
Proposed double-trap potential made of an inner trap of the form given in Eq.~\ref{InnerTrap} 
and outer trap. As outer traps we use either infinite walls at $\pm C$ (black) or a smooth power potential (red).}
%The following parameters of the trap potential (8)
%have been used: $\delta=0.1$, $A=0.04$  and $x_0=1.5$.}
\label{fig1}
\end{figure}

\begin{figure}
\includegraphics[width=11.2cm, angle=-90]{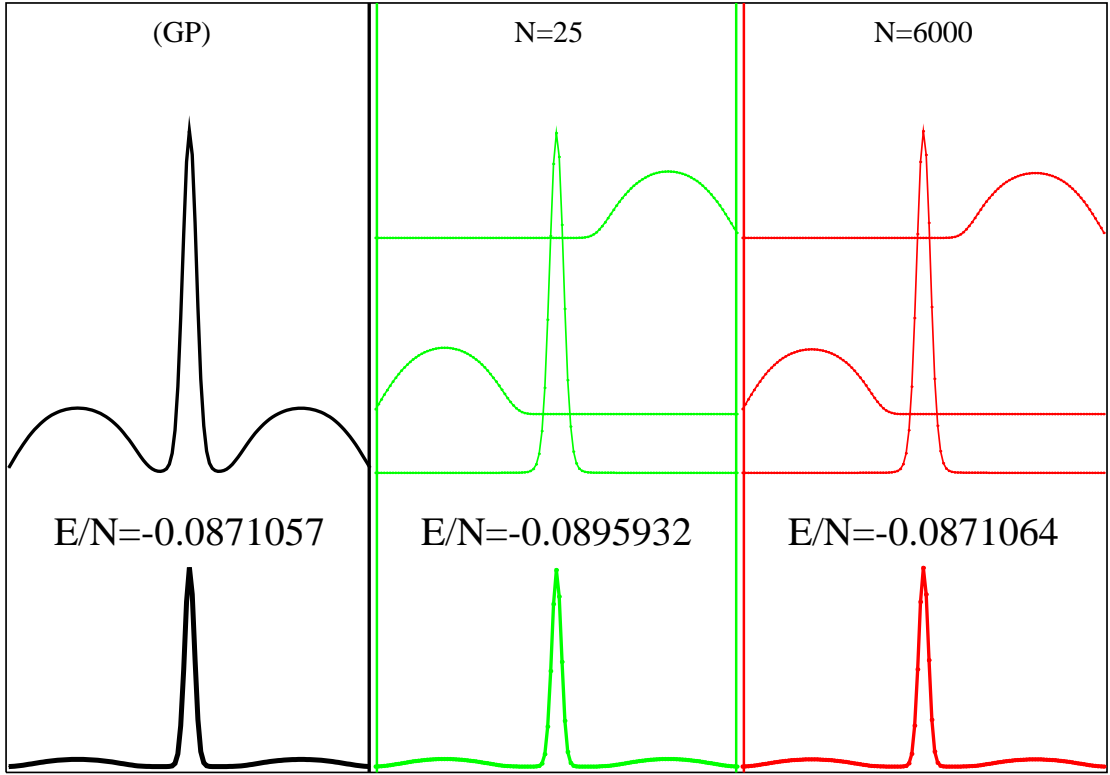}
\caption{The orbitals and densities for the double-trap potential 
with infinite walls (see black curve in Fig.\ref{fig1}).
The orbitals $\phi_1(x)$, $\phi_2(x)$ and $\phi_3(x)$ and the respective density per particle 
$\rho_{BMF}=( n_1|\phi_1|^2+ n_2|\phi_2|^2 + n_3|\phi_3|^2)/N$ of the three-fold fragmented 
ground state (for convenience the $(n_i/N)^{1/2}\phi_i(x)$ are shown) are depicted in comparison with the 
GP orbital $\varphi$ and its density $\rho_{GP}=|\varphi|^2$ for $N=25$ and $N=6000$ and $\lambda=1.3$. 
The energy per particle is indicated. For convenience, the base-line of the orbitals $\phi_2(x)$ and $\phi_3(x)$
has been moved upwards artificially from zero.
}
\label{fig2}
\end{figure}

\begin{figure}
\includegraphics[width=11.2cm, angle=-90]{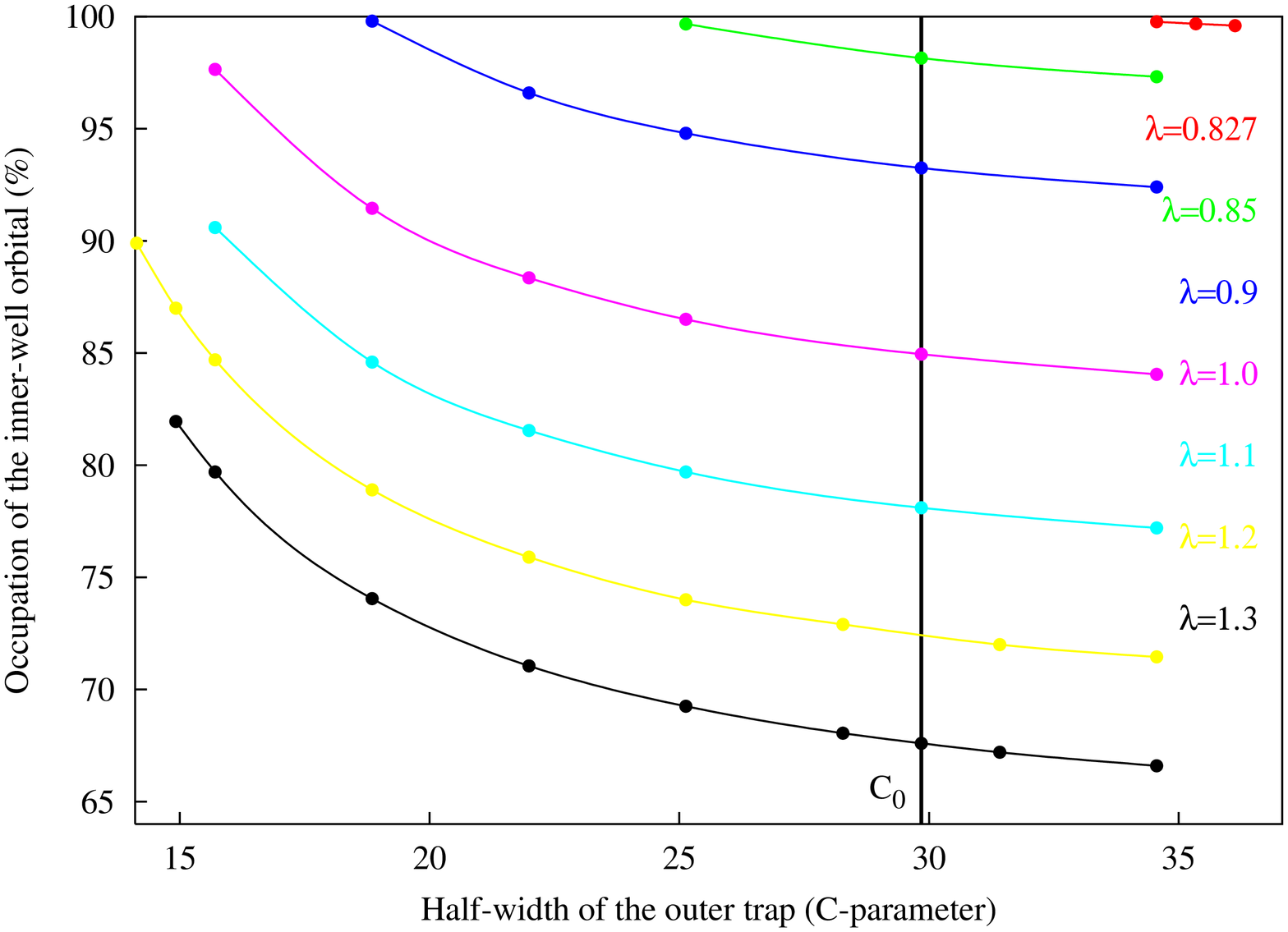}
\caption{ Manipulating the fragmentation by varying the outer trap.
Shown is the fractional occupation of the orbital localized in the inner well as a function of $C$ 
(half-width of the outer trap).
All other parameters are kept at their reference values $A_0$ and $B_0$.
}
\label{fig3}
\end{figure}

%=================== Critical parameters of Fragmentation ========================================== 
Figure \ref{fig3} shows the fractional occupation of the orbital localized in the central well as a function of
the positions $\pm C$ of the outer walls (see Figure \ref{fig1}) for several values of $\lambda$. 
From this figure it is clear that the fragmentation starts to take place
when $\lambda$ exceeds some threshold (for the example of 25 bosons and $C=11\pi$ $\lambda_{cr}=0.8249$).
It is interesting to notice that the exact value of this threshold for a system with 
a finite number of bosons is slightly smaller than that for the open system, i.e. $C\to\infty$. 
%very large N \cite{FiniteN1-2-3foldfrag}.
As N grows, the critical value of $\lambda$ obtained for the closed system
approaches the numerical result $\lambda_{cr}=0.8279$ for the open one 
where tunneling through the barriers begins  \cite{Nimrod}. 

%========================= Other limit of very big Lambda =============================
Further increasing the bosons interaction strength $\lambda$ beyond $\lambda_{cr}$ 
(at least up to $\lambda=3.0$) leads to a more pronounced fragmentation of the  ground state.
Here, we have to mention that there is another limit where fragmentation must disappear, namely 
when $\lambda$ becomes so large that the chemical potential is larger than the barrier heights of the
inner trap and particles can flow freely into the outer trap.

These observations reveal that fragmentation of the ground state is a critical phenomenon initiated 
when the number of bosons exceeds some critical number, for a fixed scattering length,
or at some critical scattering length if the number of bosons is fixed.

%=================== Move walls to(out)wards enhance or suppress fragmentation
Figure \ref{fig3} also illustrates that fragmentation can be observed if 
the width of the outer trap is quite large or more specifically, 
if the infinite walls are placed at comperativelly large distances from the barriers.
For example, for the chosen shape of the inner trap ($A_0,B_0$) with the barriers at $x_b\simeq\pm3.4$, 
and $\lambda=1.3$, about 
$30\%$ of bosons are shared by outer wells if the walls are at $C\approx25$.
By pushing the walls toward the barriers, the fragmentation gradually decreases: for $C\approx16$ 
the fragmentation is $20\%$, and for walls at $C<10$ the fragmentation totally disappears - 
all bosons are accumulated in the inner trap.
Therefore, by pushing the walls toward the barriers the fragmentation can be reduced or even totally suppressed. 
These results show that fragmentation in the double-trap potential 
may be manipulated by the outer trap:
{\it fragmentation can be suppressed by squeezing the outer trap and enhanced by its expansion.}

On the other hand, when the walls are moving outward from the barriers, the fragmented fraction
in the outer wells becomes larger and converges to some constant value which depends on $\lambda$, of course.
This value can be extracted from the results of Fig.\ref{fig3} and also
from the results obtained for the open system;
an open system may be thought of as a closed one with walls placed at infinity.
From this we conclude that for a given value of $\lambda$, 
the maximal fraction of the bosons in the outer wells 
is defined by the inner-trap potential only. 

\begin{figure}
\includegraphics[width=11.2cm, angle=-90]{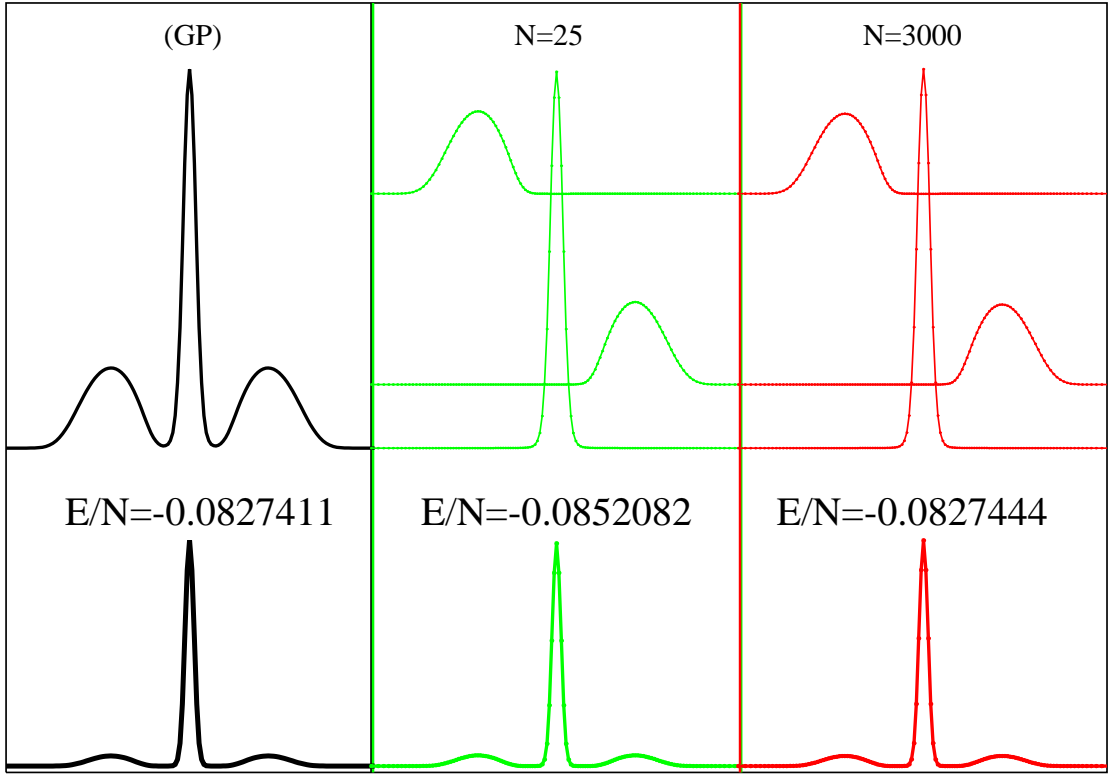}
\caption{The orbitals and densities for the double-trap potential 
with a smooth power outer trap (see red curve in Fig.\ref{fig1}).
The orbitals $\phi_1(x)$, $\phi_2(x)$ and $\phi_3(x)$ and the respective density per particle 
$\rho_{BMF}=( n_1|\phi_1|^2+ n_2|\phi_2|^2 + n_3|\phi_3|^2)/N$ of the three-fold fragmented 
ground state (for convenience the $(n_i/N)^{1/2}\phi_i(x)$ are shown) are depicted in comparison with the 
GP orbital $\varphi$ and its density $\rho_{GP}=|\varphi|^2$ for $N=25$ and $N=3000$ and $\lambda=1.3$. 
The energy per particle is indicated. For convenience, the base-line of the orbitals $\phi_2(x)$ and $\phi_3(x)$
has been moved upwards artificially from zero.
}
\label{fig4}
\end{figure}

The relevant factor for the fragmentation is the width of the outer trap
and we may suppose that the specific shape of the outer trap is of lesser importance.
To support this expectation, we study the fragmentation for  $\lambda=1.3$ in the   
systems of $N=3000$ and $N=25$ bosons trapped in the double-trap,
with the same inner potential as before and 
the smooth power $(0.035x)^{10}$ outer potential (red solid line in Fig.~\ref{fig1}).
As seen in Fig.\ref{fig4}, the three-fold fragmentation of the ground state is again 
favorable energetically for these systems.
Moreover, the optimal occupations of the inner orbital for $N=3000$ and $N=25$  
are $n_1/N\approx 72.2\%$ and $n_1/N\approx 71\%$ respectively, and hence similar to those
discussed above for the infinite walls case.
In Fig.~\ref{fig4} we plotted the orbitals and the spatial densities
determined for the smooth power outer trap.
By comparing these orbitals with those shown in Fig.\ref{fig2} for the infinite square outer trap, 
we conclude that the shape of the orbital localized in the inner trap 
does not depend upon the specific shape of the outer potential.
The profiles of the orbitals localized in the outer wells exhibit differences
which may be experimentally observed.
In the case of the smooth power outer trap the density profiles of the outer orbitals 
are gaussian-like, whereas those for the infinite square outer trap are of a sinusoidal type.
%Very similar orbital's occupation numbers may be obtained for double-trap potential
%with infinite squared outer well if the walls are placed at ($C~=\pm21$).
%%The differences in the shape of the outer orbitals may be observed when
%the relative occupations of the left and right outer wells are different.
%%Such a situation can be achieved if the center of the outer potential is displaced 
%with respect to the inner one.
%This plot gives an idea how 

\section{Manipulating Fragmentation by varying the inner trap}
\begin{figure}
\includegraphics[width=11.2cm, angle=-90]{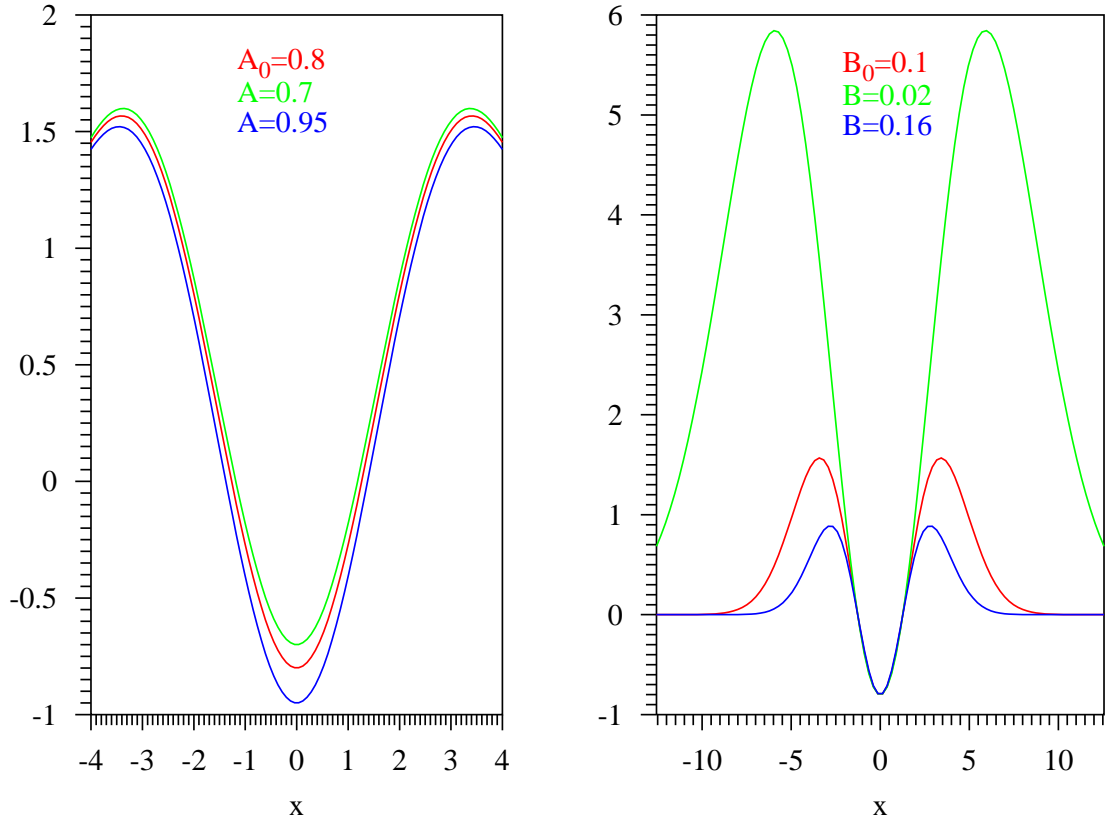}
\caption{ Parameterization of the inner trap.
Left figure: Dependence on the parameter $A$.
Right figure: Dependence on the parameter $B$.}
\label{fig5}
\end{figure}

%As we found above, the maximal fraction of the bosons which 
%can be shared by the outer walls for a fixed $\lambda$ is governed by the inner trap geometry.
%The outer trap allows to manipulate it. 
Here, we investigate how fragmentation of BEC depends upon the shape of the inner potential.
This potential permits two degrees of manipulation
by varying the depth of the inner well and by varying the height of the barriers.
The parameter $A$ of the inner potential (see Eq.\ref{InnerTrap} and Fig.\ref{fig5}A) is directly related to the 
depth of the inner well. This depth grows as $A$ is increased.
In Fig.~\ref{fig5}B we illustrate the potential dependence upon $B$, a parameter which 
defines the height of the barriers, their widths, and the positions of their maxima. 
By decreasing the value of $B$, the height of the barriers and their widths grow while
the positions of their maxima are shifted outward.

\begin{figure}
\includegraphics[width=11.2cm, angle=-90]{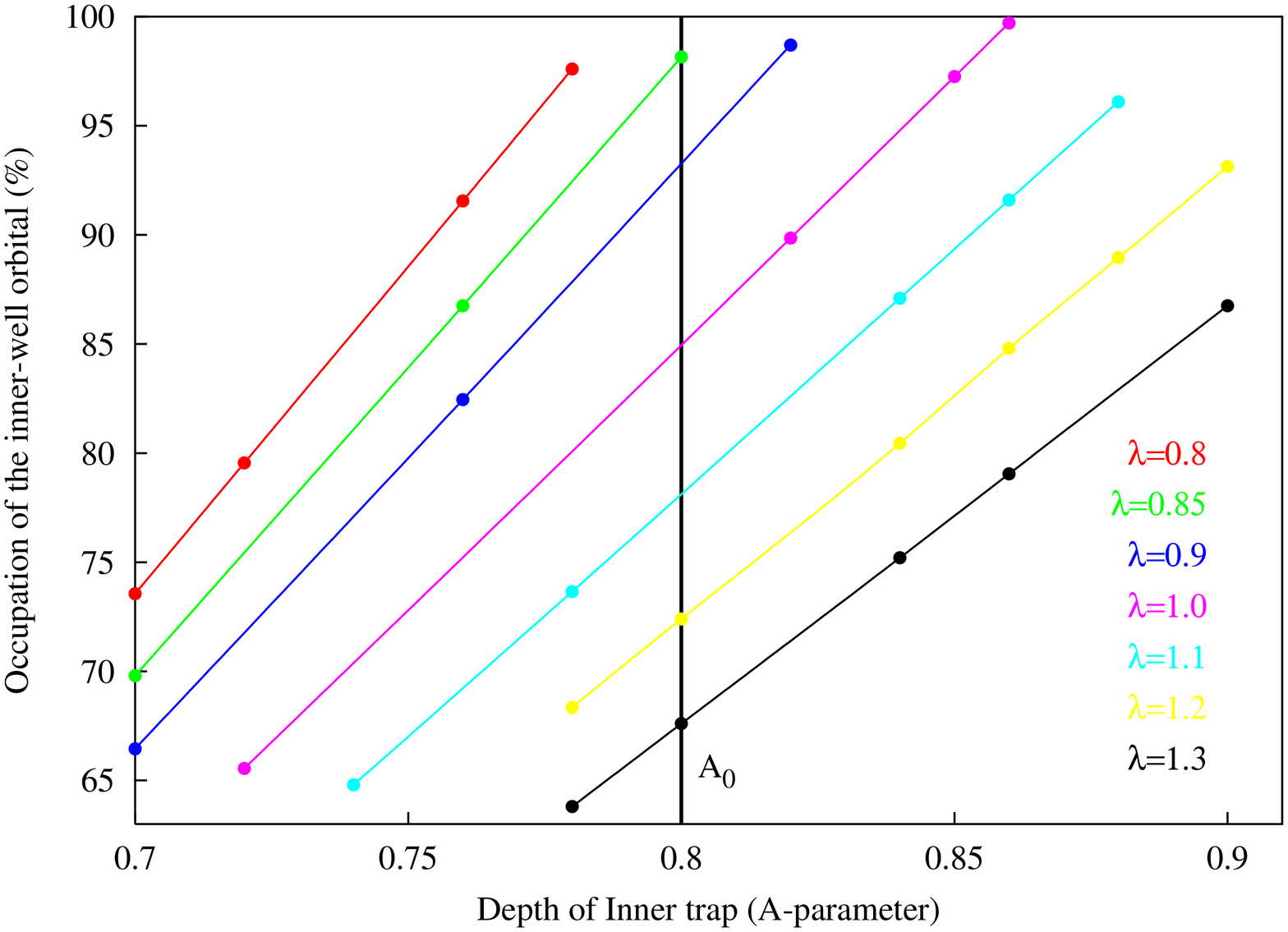}
\caption{ Manipulating the fragmentation by varying the inner trap.
Shown is the fractional occupation of the orbital 
localized in the inner well as a function of $A$.
All other parameters are kept at their reference values $B_0$ and $C_0$.}
\label{fig6}
\end{figure}

In Fig.\ref{fig6} we plot the fractional occupation in the inner trap 
as a function of $A$ for several values of $\lambda$.
It is seen that fragmentation decreases monotonically with $A$. 
Qualitatively speaking, this finding implies that by increasing the depth of the inner potential
more "room" becomes available for bosons in the inner well. Conversely, by decreasing $A$ the capacity 
of the inner well becomes smaller and more particles "flow" out into the outer wells.
This picture serves also as a verification and extension of the conclusions drawn above.
We have established that at fixed trap geometries fragmentation takes place at 
$\lambda\approx\lambda_{cr}$ and becomes more pronounced as $\lambda$ is increased.
We may now add that there is a critical depth of the inner potential 
initiating fragmentation. For example, for $\lambda=0.8$ and $A=0.8$ ($B_0=0.01,C_0=9.5\pi$)
fragmentation does not exist, while for $A=0.72$ the system becomes $20\%$ fragmented.

\begin{figure}
\includegraphics[width=11.2cm, angle=-90]{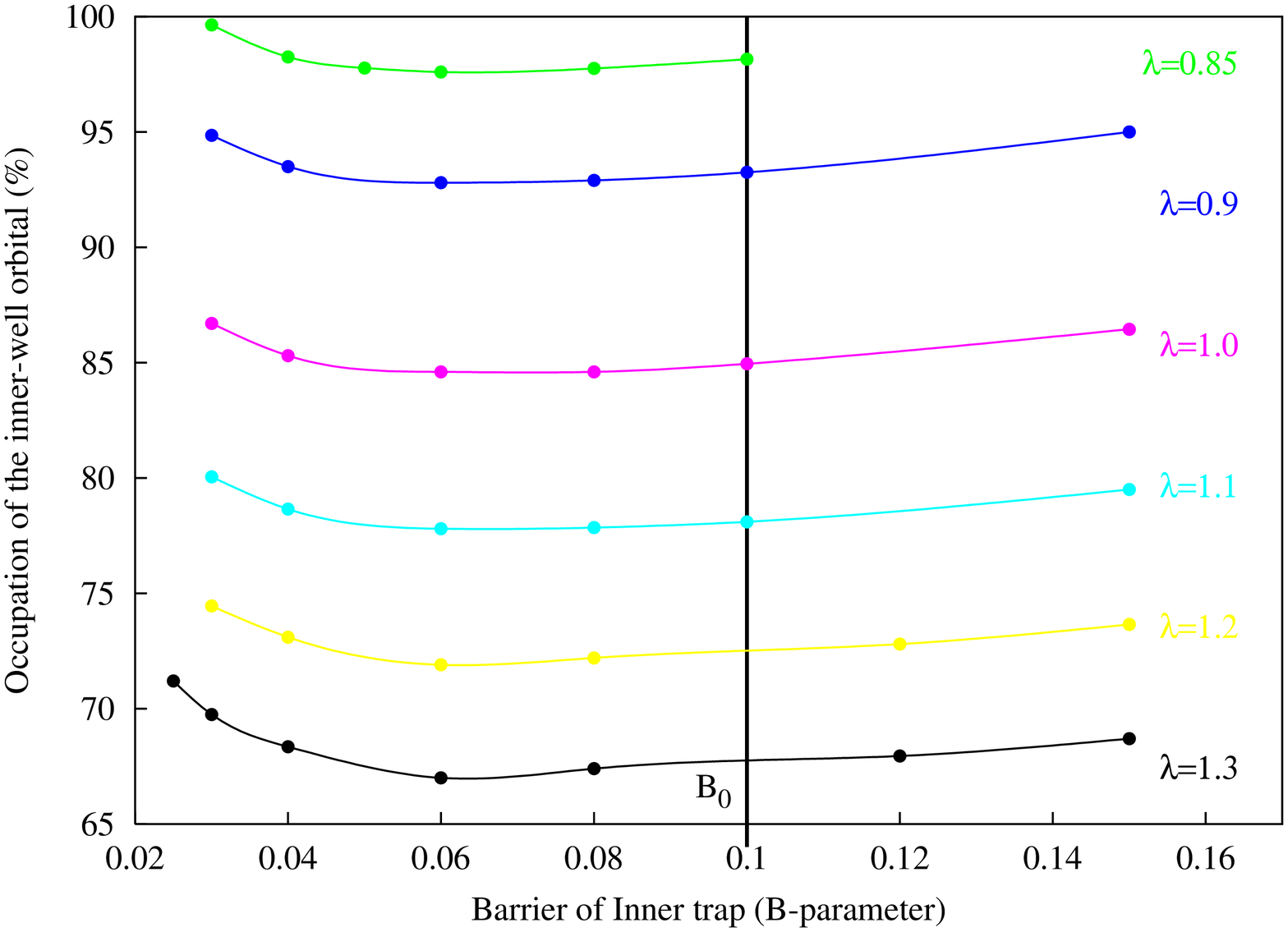}
\caption{ Manipulating the fragmentation by varying the inner trap.
Shown is the fractional occupation of the orbital 
localized in the inner well as a function of $B$.
All other parameters are kept at their reference values $A_0$ and $C_0$.}
\label{fig7}
\end{figure}

Finally,  we investigated the dependence of the fragmentation on 
the height of the barriers ($B$-parameter).
In Fig.~\ref{fig7}, we plot the fractional occupation of the inner well as a 
function of $B$ ($A_0=0.8,C_0=9.5\pi$) for several $\lambda>\lambda_{cr}$.
From this figure it is clear that as long as $B$ is not too small or too large the fragmentation 
is not particularly sensitive to variations of $B$.
%because the strongly increasing the barrier's height results in small changes of the occupation numbers.
Decreasing B from 0.16 to 0.02 corresponds to a substantial change of the barriers heights from 0.8 to 5.8 units. 
At the same time the respective fractional occupation of the bosons in 
the central well varies by several percents only for any fixed value of $\lambda$.
A further decrease of B causes, however, the disappearance of the fragmentation.
Such a behavior is to be expected in this case, since a very small value of $B$ corresponds to very broad 
barriers (see Fig.\ref{fig5}) diminishing thereby the size of the outer wells and hence
their capacity to hold bosons at favorable energy cost (see. Sec.V).
%The fragmentation is also dissappeared  in the other limit 

On the other hand, at large values of $B$ the height of the barriers becomes very small and
the fragmentation of the ground state is expected to be unfavorable energetically.
This issue is further discussed in the subsequent section.
%Indeed, the three-fold fragmentation of the ground state 
%for the systems of $N=25$ and $N=5\cdot10^7$ bosons at $\lambda=2.5$
%is found to be energetically unfavorable. 

It is very important to note that the presence of barriers is an essential factor 
for the ground state fragmentation of the repulsive BEC. 
The specific shape of the barriers is of lesser relevance and its impact on fragmentation
can be largely compensated by moving the position of the walls of the outer trap.
We expect, however, that for time-dependent studies this situation will be changed drastically, because 
the tunneling time (i.e. the time which is needed to tunnel through the barriers) is determined by 
the height and width of the barriers.

\section{Manipulating the maximal number of bosons in the fragmented ground state}
There are several factors limiting the number of bosons in a fragmented ground state.
The most transparent factor is the height of the barriers of the inner trap.
Increasing the particle number N obviously enlarges $\lambda=\lambda_0(N-1)$
for a given scattering length. We have already pointed out that when $\lambda$ becomes so large that
the chemical potential is larger than the barrier heights, particles can flow freely out of the inner trap
and the fragmentation disappears. Consequently, there is a maximal number
$N_{max}$ of bosons in a fragmented ground state and this number depends on the double-trap potential and
on the scattering length.

For the open system we have seen that- at a fixed value of $\lambda$- once fragmentation in the
ground state takes place, i.e., $\lambda>\lambda_{cr}$, this fragmentation persists for
any number of bosons (N larger than 1, of course). This finding does not hold for
closed systems. For these systems there is again an $N_{max}$ even if $\lambda$ is kept 
fixed at a value where the chemical potential is smaller than the barriers heights.
As seen in the preceding sections, there is an enormous range of double-trap potentials
giving rise to fragmented ground states as long as $\lambda>\lambda_{cr}$.
The degree of fragmentation can be widely manipulated by varying the parameters of the trap 
potentials. Two major questions arise now: what is the origin of $N_{max}$ and can this value
be manipulated by varying the parameters of the trap potentials? The answer to the latter question
is positive and we shall present numerical calculations below.

To facilitate the origin of the existence of $N_{max}$ we remind that $\lambda=\lambda_0(N-1)$
where $\lambda_0$ is the interaction strength of two interacting bosons. Increasing the number N
of bosons while keeping the value of $\lambda$ fixed obviously implies a weakening
of the interaction strength. As N approaches infinity, the
interaction strength $\lambda_0$ approaches zero and we may expect that the ground state takes on
the appearance of a non-interacting system, i.e., that of the GP ansatz. Indeed, our numerical
calculations show that while for $N \le N_{max}$ the ground state is fragmented, fragmentation still persists even
for $N>N_{max}$, but the state in question is an excited state of the condensate.
This excited state is very low lying for {\it all} $N > N_{max}$; its energy is extremely close to that
of the now unfragmented ground state (in the present examples the two energies per particle differ just
at the $7^{th}$ digit!). The existence of a very low-lying fragmented excited state is of 
great interest by itself and should play a role in particular in time-dependent experiments.

\begin{figure}
\includegraphics[width=11.2cm, angle=-90]{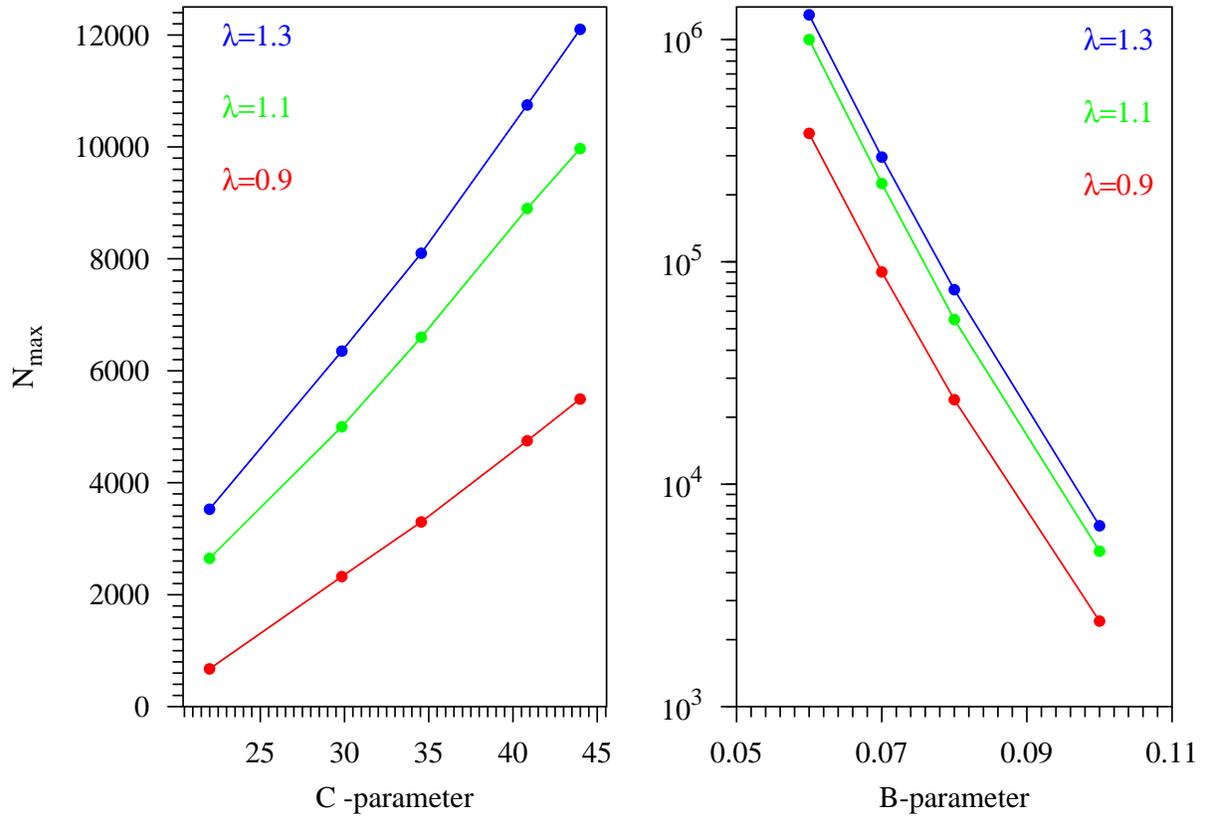}
\caption{Manipulating the maximal number of bosons $N_{max}$ in the fragmented ground state
by varying the double-trap potential.
Left figure: Dependence on the parameter $C$.
All other parameters are kept at their reference values $A_0$ and $B_0$.
Right figure: Dependence on the parameter $B$ (note the logarithmic scale).
All other parameters are kept at their reference values $A_0$ and $C_0$.}

\label{figNmax}
\end{figure}

In Fig.\ref{figNmax} we show $N_{max}$ as a function of the outer trap parameter $C$ for several 
values of $\lambda$ keeping the parameters of the inner trap at their reference values. 
Also shown are the variations of $N_{max}$ with the inner trap parameter $B$. 
In the range of parameters studied, the maximal number of bosons 
$N_{max}$ grows linearly with the size of the outer trap and 
exponentially with the height and width of the barriers.
%??  Note that a small value of the parameter $B$ reduces 
%the effective size of the outer wells; see Fig.\ref{fig5}B.  ??

For a better understanding of the above findings we draw attention to Fig.\ref{fig2}.
To maintain orthogonality, the three orbitals of the BMF live in different regions of space. 
%The geometry of the double-trap and repulsive character of the interparticle interaction
%ensure this requirement. Firstly, 
As long as the size of the outer traps is large enough, this is 
favored by the presence of broad and high barriers between the inner and outer wells.
Since the condensate is repulsive, a large outer trap enables the wavefunction in this region of space
to delocalize without the necessity to penetrate the inner well.

%As long as the size of the outer traps is large enough, this is favored by the 
%presence of broad and high barriers between the inner and outer wells.
%Since the condensate is repulsive, a large outer trap enables the wavefunction in this region of space
%to delocalize without the necessity to penetrate the inner well.

\section{Manipulation of Fragmentation by several inner traps}
\begin{figure}
\includegraphics[width=11.2cm, angle=-90]{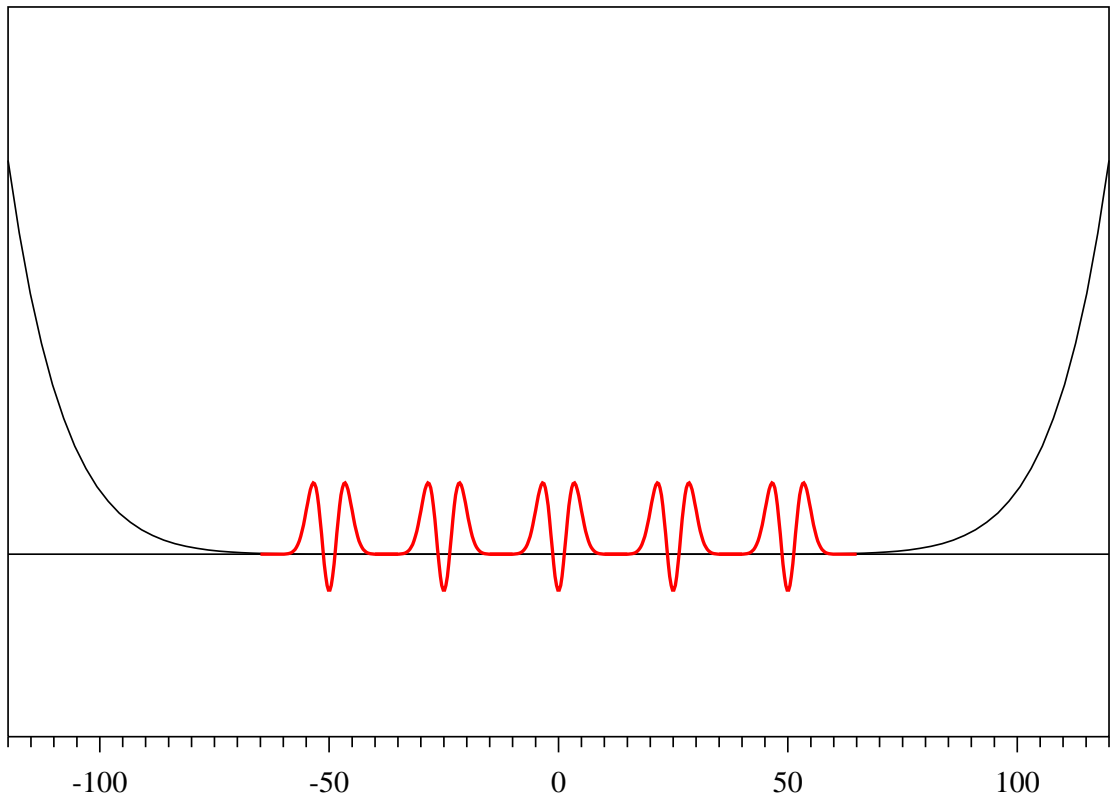}
\caption{ Proposed multi-well double-trap potential.}
\label{fig8}
\end{figure}

Three-fold fragmentation of the ground state is found to take place in suitable three-well potentials.
The results obtained for these potentials can naturally be extended to an array of multiple wells. 
This extension is particularly relevant due to the recent experimental efforts to investigate 
one-dimensional optical lattices \cite{OpttrapExper}.
%Another opportunity to manipulate the fragmentation is to use an inner potential made of several
%wells and barriers. 
For example, a multi-well potential may be formed 
if the inner-well trap given by Eq.\ref{InnerTrap} is translated several times and then embraced by an
outer smooth power or infinite square trap.
In these cases we expect many-fold fragmentation and if the inner traps are well separated 
from each other and from the outer walls, then
the critical value of $\lambda$ scales according to the number of inner potentials.
In Fig.\ref{fig8} we plotted an example of such a trap.
The critical value of $\lambda$ for this system is approximately five times larger than the respective value
obtained above for the single inner potential. 
This value as well as the extend of the fragmentation can be sensitively manipulated by varying
the trap parameters.
Of course, a many-orbital BMF must be used in order to correctly describe the situation.
For the multi-well trap shown in Fig.\ref{fig8} we expect that 11 orbitals should
be used and that should be possible to do in the future.

%The main conclusion of this observation is:
%if for some reasons the number of bosons and their scattering length is fixed,
%the many-fold fragmentation can be enforced by using the multi-well inner trap.

\section{Conclusions}
In this article we have investigated the fragmentation phenomenon in the ground state
of a repulsive condensate immersed into the double-trap potential.
We demonstrate that fragmentation can be successfully characterized by the best mean-field approach.
To be able to correctly describe $m$-fold fragmentation, $m$ orbital are needed and are available in
the $m$-orbital mean-field  BMF($m$) method.
In this method the occupation number of each fragment as well as the optimal shape of the respective 
orbitals are determined variationally by minimizing the total energy functional.
If more orbitals are included than needed, the occupation of the superfluous orbitals becomes zero
by minimizing the total energy.

The double-trap potential studied here has three wells, obtained as the superposition of 
an inner trap exhibiting one well and two barriers and an outer trap embedding the inner one.
For many choices of the potentials, the macroscopic occupation of the three orbitals 
may become energetically more favorable than accumulating all the particles in a single orbital.
The fragmentation of the ground state is found to occur  
when the number of bosons exceeds some critical value which depends on the scattering length and
on the shape of the inner trap potential.
For the example studied we found that if fragmentation is observed for a large number of bosons,
then it exists also for any smaller number of bosons (of course $N>1$) when $\lambda$ is kept fixed.
When $\lambda$ is kept fixed, there exists, however, a maximal number of bosons
for which the ground state is fragmented. This number can be strongly manipulated by varying the 
double-trap potential.
%The relative fragmentations in both cases may differ in order of several percents only.

We have demonstrated that the geometry of the inner potential 
determines the values of the critical parameters. Moreover, for any given number of
bosons this potential also determines the maximally possible fragmented fraction of bosons 
which is localized in the outer wells.
The actual fragmented fraction of bosons may be effectively manipulated by the proper choice of the outer trap.
The interplay between the inner and outer trap potentials provides a sensitive tool to
manipulate fragmentation of repulsive condensates. Varying the number of bosons in the condensate and the
scattering length are also instrumental in this respect.

The results obtained for three-well potentials can naturally be extended to an array of multiple wells. 
%In particular, three-fold fragmentation is predicted if as an outer trap the infinite-square and harmonic wells
%are to be chosen. 
%By displacing the infinite-square or harmonic inner trap with respect to the outer one 
%the relative occupations of the outer wells may be changed. 

%The natural extensions of mentioned above logic is to consider such a trap potential 
%or potentials where several competing processes are available.
%The main process is energetical - the energy spleet between lowest energy levels, - an 
%ideal situation for succesfull realization (completion) in this case is symmetric 
%double(multi) well potential with a barrier between wells. For repulsive condensates it has been shown, 
%however, that single-orbital ground state is energetically more favorable then fragmented ones.
%This result can be explained (understood) by the fact that
%the density profiles for both orbitals are almost the same.
%Therefore, in order to enforce fragmentation one needs to consider such a multi-well potential where the densities
%of the ground and excited states are totally different.

The authors acknowledge useful discussions with Ofir Alon and Kaspar Sakmann.

%\pagebreak

\pagebreak
%Insert fig1.1

%%Insert fig2

%%Insert fig3

%%Insert fig4

%%Insert fig5

%%Insert fig6

%%Insert fig7

%%Insert fig N_max

%%Insert fig8

\end{document}